\documentclass[10pt,a4paper]{article}
\usepackage{amsmath,amssymb}
\usepackage[T1]{fontenc}
\usepackage[utf8]{inputenc}

\newtheorem{satz}{Theorem}[section]

\newtheorem{assumption}[satz]{Assumption}
\newtheorem{defi}[satz]{Definition}

\newtheorem{bem}[satz]{Remark}
\newtheorem{lemma}[satz]{Lemma}
\newtheorem{koro}[satz]{Corollary}
\newtheorem{conclusion}[satz]{Conclusion}
\newtheorem{ob}[satz]{Observation}

\newtheorem{propo}[satz]{Proposition}
\newtheorem{postulate}[satz]{Postulate}
\newtheorem{conjecture}[satz]{Conjecture}

\newcommand{\mcal}{\mathcal}
\newcommand{\mbf}{\mathbf}
\newcommand{\mbb}{\mathbb}
\newcommand{\tit}{\textit}

\newcommand{\beq}{\begin{equation}}
\newcommand{\eeq}{\end{equation}}

\begin{document}
\thispagestyle{empty}
\begin{center}
\vspace*{1.0cm}
{\Large{\bf The Crossed Product, Modular (Tomita)\\ Dynamics and its Role in the Transition of \\Type $III$ to Type $II_{\infty}$ v.Neumann\\ Algebras and Connections to Quantum Gravity}}
\vskip 1.5cm
 
{\large{\bf Manfred Requardt}}

\vskip 0.5cm

Institut fuer Theoretische Physik\\
Universitaet Goettingen\\
Friedrich-Hund-Platz 1\\
37077 Goettingen \quad Germany\\
(E-mail: muw.requardt@googlemail.com) 

\end{center}
\begin{abstract}
We analyse the role of the crossed product and the modular (Tomita) dynamics in the transition of type $III$ to type $II_{\infty}$ v.Neumann algebras which was recently observed in papers by Witten et al. In a preceding paper we argued that type $II_{\infty}$ v.Neumann algebras display certain features which we attributed to quantum gravity effects. We claim that the action of the modular evolution on the quantum fluctuations can be understood as an aspect of quantum gravity. We mention in this context the work of Sakharov on induced gravity. Furthermore we analyse the change of the properties of projectors and partial isometries in the transition from type $III$ to type $II_{\infty}$.
\end{abstract}
\newpage
\section{Introduction}
The \tit{crossed product} plays an important technical role in the (type) analysis of v.Neumann algebras, in particular the \tit{duality result} of M.Takesaki \cite{Take}. This type analysis was initiated by Murray and v.Neumann in the thirties/fourties of the 20th century (see e.g. \cite{N1},\cite{N2},\cite{N3},\cite{N4} or \cite{N5}). Modern treatments can be found in a number of more recent books (cf. the references in \cite{type}).

As in the following we mainly deal with properties and implications of the crossed product, we mention rather a few sources where it is treated and discussed in more detail, see for example \cite{Take2},\cite{Daele} or \cite{Kad2}. In the following we will rely mainly on the notations of the comprehensive discussion in \cite{Kad2}.

In mathematical physics v. Neumann algebras and their properties played an important role in algebraic quantum field theory and quantum statistical mechanics (see for example \cite{Brat1},\cite{Brat2},\cite{Haag}. More recently their type theory and the role of the crossed product were studied and exploited in various applications in black hole physics, holography and some other subfields where quantum field theory (QFT) and gravity meet and interact (\cite{W1},\cite{W2},\cite{W3},\cite{W4},\cite{Ahmad1},\cite{Ahmad2},\cite{Ahmad3}). A crucial role in this context is played by the concept of entanglement entropy of (neighboring) regions in curved space time (CST), which is notoriously uv-divergent in QFT.

Typically, local observable algebras are of type III, having the consequence that no finite or semifinite trace does exist. On the other hand, type $II_{\infty}$ v. Neumann algebras have a semifinite trace and type $II_1$ algebras even have a finite trace. Therefore, in these cases a (generalized) entropy can be defined which is considered to be of great relevance. It is a fundamental property of the crossed product that it transforms type III into type II, while another application transforms type II back into type III. This is the content of the fundamental duality result of Takesaki et al. Another recent application of the crossed product in in the field of socalled \tit{quantum reference frames} (see e.g. \cite{Ahmad4} or \cite{Fewster} for a very incomplete list). An older but related paper is \cite{Ojima}.

In the following we want to ask some fundamental questions of principle. We will concentrate ourselves on the role of the Tomita modular action as automorphism group in the crossed product. As argued in e.g. \cite{Connes},\cite{Rov1},\cite{Rov2},\cite{Dirac} we expect that the quasi automatic existence of this symmetry group in the context of type III v. Neumann algebras point to a fundamental role it is playing in the relation of QFT and gravity. As the role of the Tomita action is in most cases not directly observable via, for example, measurements in QFT, its actual physical content is not entirely obvious and is the subject of various speculations.

It is one of our aims to relate each step in the mathematical construction of the crossed product to an ontological correlate in the physical substructure underlying this field of QFT entangled with (quantum) gravity or quantum space time (QST). At the same time we want to exhibit the existence of a preestablished harmony between the extremely rich structure of v. Neumann algebras and the laws governing QFT interacting with  (quantum) gravity.

To describe some of the steps in this process:\\
i) We interpret the role of the Tomita evolution as the effect of (quantum) gravity on the fluctuation spectrum (in particular the vacuum fluctuations) of QST and QFT in CST. Typical cases in point are the fundamental analysis done by Fulling and the corresponding description by Unruh of QFT in the \tit{Rindler wedge} (\cite{Fulling1},\cite{Fulling2},\cite{Unruh}).\\
ii) Furthermore we regard, as is also described in \cite{Connes},\cite{Rov1},\cite{Rov2},\cite{Dirac} in some detail, the Tomita modular structure of type III v. Neumann algebras as encoding the thermal structure of the quantum vacuum (cf. also \cite{Requ2},\cite{Requ1}).\\
iii) We argue that the crossed product construction combines and entangles these different (quantum) processes into one larger and extended v. Neumann algebra, thus making the largely hidden gravitational and fluctuation processes observable and measurable, as this is the physical content of being a member of a v. Neumann algebra.\\
iv) Last  but not least, we are interested in the physical meaning and implications of projectors and partial isometries as members of v. Neumann algebras of different  types, that is, type II or type III.

But before we are able to try to clarify these important points, we will in the next section give a brief summary of the various mathematical steps making up the crossed product construction and its duality property. 
\section{The Crossed Product}
In this section we provide a brief introduction into the concept of the crossed product as a tool to generate a new extended v. Neumann algebra from a given one. In \cite{type} we already discussed the various types of projectors in a v. Neumann algebra and the resulting type classification (but mainly in the context of type II v. Neumann algebras). 
\begin{bem}A nice discussion of the v. Neumann type theory can be found in \cite{Sorce}, see also \cite{types}.
\end{bem}

Before we will begin to introduce the crossed product we want to briefly recapitulate some notions concerning the type of projectors since there exist a variety of definitions which are overlapping to some extent. In the following we restrict ourselves, for convenience, to v. neumann factors.
\begin{defi}A v. Neumann factor, $\mcal{M}$, is either of type $I_n,I_{\infty},II_1,II_{\infty}$ or $III$. It is of type $I$ if it has a minimal projector, of type $I_n$ if $\mbf{1}$ is the sum of $n$ minimal projections. If $\mcal{M}$ has no minimal projection but has a non-zero finite projection, it is of type $II$, of type $II_1$ if $\mbf{1}$ is finite, of type $II_{\infty}$ if $\mbf{1}$ is infinite. If $\mcal{M}$ has no non-zero finite projection, it is of type $III$.
\end{defi}
\begin{bem}Factors of type $I_n$, $n\in\mbb{N}$ or $\infty$ are unitarily equivalent to $\mcal{B}(\mcal{H})'\otimes \mbf{1}_{P\mcal{H}}$, $\mcal{H}$ the original Hilbert space of $\mcal{M}$, $P$ a minimal projector, the dimension of $\mcal{H}'$ being the number of occurring minimal projectors in the corresponding resolution of the unit projector $\mbf{1}$. By the same token they are *-isomorphic to the algebra $\mcal{B}(\mcal{H}')$. 
\end{bem}

Another classification is:
\begin{defi}Factors of type $I_n, n\in \mbb{N}$ or $II_1$ are called finite. Factors which are not type $III$ are called semifinite. Factors of type $III$ are called purely infinite. Factors which are not type 
$I_n, n\in \mbb{N}$ or $II_1$ are called properly infinite (implying that the unit projector $\mbf{1}$ is infinite). Factors of type $I_{\infty}$ or $II_{\infty}$ are called semifinite and properly infinite.
\end{defi}
\begin{bem}It is important to note that usually the existence of a projector of a certain type in $\mcal{M}$ implies the existenceof many projectors of this type. Frequently it is even possible to represent the unit projector $\mbf{1}$ as an (infinite) sum of such mutually orthogonal and equivalent projectors.
\end{bem}
\begin{koro}In the semifinite case we can construct a (generalized) trace on $\mcal{M}$ (cf. the above mentioned literature).
\end{koro}
\begin{bem}(Historical) It is a little bit funny that since the work of v. Neumann et al. v. Neumann algebras are frequently denoted by the letter $\mcal{M}$. A reason for v. Neumann may have been that operators have been usually denoted by $A,B,C,\ldots$ and the used letter M should be sufficiently separated from these letters in the alphabet.
\end{bem} 

In the following type $III$ and type $II_{\infty}$ will become important.
\begin{defi}A v. Neumann algebra is called $\sigma$-finite or countably decomposable if sequences of mutually othogonal projectors are at most countable.
\end{defi}
\begin{bem} If the Hilbert space is separable this property holds.
\end{bem}
\begin{ob}In a type $III$ v. Neumann algebra all projectors are infinite. In particular, in the $\sigma$-finite case it holds for the unit projector $\mbf{1}$ and each projector with $P\leq \mbf{1}$ that there does exist a partial isometry $V$ with 
\beq  V:\mcal{H}\to P\mcal{H}\;\text{isometrically and}\; V^*V=\mbf{1}\;,\; VV^*=P  \eeq
that is, $P\sim\mbf{1}$ for all $P\in\mcal{M}$. This implies that all projectors in $\mcal{M}$ are equivalent.
\end{ob}

In the case $II_{\infty}$ there exist finite and infinite projectors. With $P$ finite it holds for each true subprojection $E<P$ that there does not exist a partial isometry with the above properties (i.e., mapping $E\mcal{H}\to P\mcal{H})$. Furthermore it holds:
\beq \mbf{1}\;=\;\sum_{i=1}^{\infty}\;P_i  \eeq
with all $P_i$ equivalent to a finite projector $P_1$ (cf. the discussion in \cite{type}).

We now start to introduce the (abstract) crossed product of a v. Neumann algebra $\mcal{M}$, acting on the separable Hilbert space $\mcal{H}$ and carrying an automorphic representation $\alpha(s)$ of the locally compact abelian group $\mbb{R}$ with its dual group $\hat{\mbb{R}}$. If the automorphism group is implemented unitarily, i.e.
\beq \alpha_s(A)=U_sAU_s^{-1}   \eeq
the crossed product is called the implemented crossed product.

We define the Hilbert space $L_2(\mbb{R},\mcal{H})$. It can be shown that
\begin{lemma}$L_2(\mbb{R},\mcal{H})$ is unitarily equivalent to the tensor product $\mcal{H}\otimes L_2(\mbb{R})$ with 
\beq W(u\otimes f)(s):=f(s)\cdot u  \eeq
\end{lemma}
\begin{bem}In the following we identify $L_2(\mbb{R},\mcal{H})$ with  $\mcal{H}\otimes L_2(\mbb{R})$, thus usually suppressing the operator $W$. But note that a general member of  $L_2(\mbb{R},\mcal{H})$ is frequently represented in  $\mcal{H}\otimes L_2(\mbb{R})$ by an infinite sum.
\end{bem}

With $g\in L_{\infty}(\mbb{R})$, the equation
\beq  (M_gx)(s)=g(s)\cdot x(s)  \eeq
defines a bounded linear operator, $M_g$, on $L_2(\mbb{R},\mcal{H})$. We have
\beq M_g(u\otimes f)(s)=g(s)(u\otimes f)(s)=g(s)f(s)u=(u\otimes g(s)f(s)) \eeq
hence
\beq M_g=\mbf{1}\otimes m_g  \eeq
\begin{propo}With $a(s)$ a weakly operator continuous map from $\mbb{R}$ into a bounded subset of $\mcal{B}(\mcal{H})$
\beq (Ax)(s)=a(s)x(s)\quad ,\quad x\in L_2(\mbb{R},\mcal{H})  \eeq
defines a bounded opearator on $L_2(\mbb{R},\mcal{H})$. With $\mcal{M}$ acting on $\mcal{H}$ and $a(s)\in \mcal{M}$ we have 
\beq A\in \mcal{M}\overline{\otimes}\mcal{A}\subset \mcal{M}\overline{\otimes} \mcal{B}(L_2(\mbb{R}))  \eeq
with $\mcal{A}$ the multiplication algebra $\{m_g\; ,\; g\in L_{\infty}(\mbb{R})\}$ acting on $L_2(\mbb{R})$.
\end{propo}
The proof can be found in \cite{Kad2}. A starting point is the following simple calculation. For $a(s)=g(s)A_0\sim A_0\otimes g$ and $x(s)=f(s)u\sim u\otimes f$ we have
\begin{multline}
A\cdot(gA_0)(s)=A((u\otimes f))(s)=a(s)(u\otimes f)(s)= g(s)f(s)A_0u \\
(A_0u\otimes gf)(s)=(A_0u\otimes m_gf)(s)=(A_0\otimes m_g)(u\otimes f)
\end{multline}
\begin{bem}While $A$ is lying in $\mcal{M}\overline{\otimes} \mcal{A}$ one should note that usually one needs infinitely many terms to represent it.
\end{bem}

For each $A\in \mcal{M}$ the mapping $s\to\alpha_s^{-1}(A)$ is bounded and weakly operator continuous. By the above proposition 
\beq (\Psi (A)x)(s):=\alpha_s^{-1}(A)x(s)\quad ,\quad x\in  L_2(\mbb{R},\mcal{H})  \eeq
defines a bounded linear operator, $\Psi (A)$, acting on $L_2(\mbb{R},\mcal{H}) \sim \mcal{H}\otimes L_2(\mbb{R})$ and
\beq \Psi (A)\in \mcal{M}\overline{\otimes}\mcal{A}  \eeq
\begin{koro}$\Psi$ is a $\star$-isomorphism of $\mcal{M}$ on a v. Neumann subalgebra of $\mcal{M}\overline{\otimes} \mcal{A}$.
\end{koro}

We can define the following unitary operators:
\begin{propo}
For all real $t,p$ we can define
\begin{align}
 (l_tf)(s):=f(t-s)\quad &,\quad L_t(t)(s):=x(s-t) \\
 (w_pf)(s):=\exp(-isp)f(s)\quad &,\quad (W_px)(s):=\exp(-isp)x(s)
 \end{align}
 $l_t,w_p$ are unitary operators on $L_2(\mbb{R})$, $L_t,W_p$ are unitary operators on $L_2(\mbb{R},\mcal{H})$ (or $\mcal{H}\otimes L_2(\mbb{R})$.
\end{propo} 
We have
\beq L_t=\mbf{1}\otimes l_t\quad ,\quad W_p=\mbf{1}\otimes w_p \eeq
and the following commutation relations hold
\begin{align}
w_pl_tw_p^*=\exp(-itp)l_t \quad &,\quad L_t\Psi(A)L_t^*=\Psi(\alpha_t(A))  \\
W_p\Psi(A)W_p^*=\Psi(A)\quad &,\quad W_pL_tW_p^*=\exp(-itp)L_t
\end{align}
\begin{bem}Obviously $l_t$ is a translation operator. Hence $\hat{p}=i^{-1}d/ds$ is the infinitesimal generator. In physics it is called momentum operator.
\end{bem}
 
We now are ready to introduce the (abstract) crossed product of $\mcal{M}$ and the automorphism group $\alpha(s)$. In a first step we show the following:
\begin{multline}
(\Psi(A_1)\cdot L_{t_1})\cdot (\Psi (A_2)\cdot L_{t_2})=\Psi(A_1)(L_{t_1}\Psi(A_2)L_{t_1}^*)L_{t_1}\cdot L_{t_2} \\
=\Psi(A_1)\cdot \Psi(\alpha_{t_1}(A_2))\cdot L_{t_1+t_2}=\Psi(A_1\cdot \alpha_{t_1}(A_2))\cdot L_{t_1+t_2}
\end{multline}
That is, terms like $(\Psi(A)L_t)$ or $(L_t\Psi(A))$ and their finite sums  define a $\star$-subalgebra $\mcal{R}_0$ of $\mcal{M}\overline{\otimes}\mcal{B}(L_2(\mbb{R}))$ with $W_p\mcal{R}_0W_p^*=\mcal{R}_0$ and we define:
\begin{defi}The weak closure of $\mcal{R}_0$, i.e. the v. Neumann algebra generated by the set
\beq \{\Psi(A),L_t;A\in \mcal{M},t\in \mbb{R}\}  \eeq
is called the (abstract) crossed product, $\mcal{R}(\mcal{M},\alpha (s))$ acting on 
\beq L_2(\mbb{R},\mcal{H})=\mcal{H}\otimes L_2(\mbb{R})  \eeq
\end{defi}
The dual representation of the automorphism group $\alpha(s)$ on $\mcal{R}(\mcal{M},\alpha (s))$
is $\hat{\alpha}_p$ implemented by the unitary operators $W_p$ (for the dual group of group characters see e.g. \cite{Rudin}).

If $\alpha_s$ is implemented by $U_sAU_s^*$, $U$ on $L_2(\mbb{R},\mcal{H})$ defined by $(Ux)(s)=U_sx_s$ we have:
\begin{align}
(U\Psi(A)U^*x)(s)&=((A\otimes\mbf{1})x)(s)  \\
(UL_tU^*x)(s)&=((U_t\otimes l_t)x)(s)  \\
(UM_gU^*x)(s)&=(M_gx)(s)  \\
\end{align}
That is:
\beq U\Psi(A)U^*=A\otimes\mbf{1}\,,\,UL_tU^*=U_t\otimes l_t\,,\,U(\mbf{1}\otimes m_g)U^*=\mbf{1}\otimes m_g  \eeq
the rhs being defined on $\mcal{H}\otimes L_2(\mbb{R})$ (Note that e.g. $(U\Psi(A)U^*x)(s)=Ax(s)$)    . It follows that $U\mcal{R}(\mcal{M},\alpha)U^*$ is a v. Neumann algebra, generated by 
\beq A\otimes\mbf{1}\; ,\; U_t\otimes l_t\;\text{acting on}\; \mcal{H}\otimes L_2(\mbb{R})  \eeq
\begin{defi}$U\mcal{R}(\mcal{M},\alpha)U^*$ is called the implemented crossed product.
\end{defi} 
 In the same sense as for the abstract crossed product we have the corresponding product rules:
\beq (A_1\otimes\mbf{1})(U_{t_1}\otimes l_{t_1})=(A_1U_{t_1}\otimes l_{t_1})  \eeq
and
\begin{multline}
(A_1\otimes\mbf{1})(U_{t_1}\otimes l_{t_1})\cdot (A_2\otimes\mbf{1})(U_{t_2}\otimes l_{t_2}) \\
=A_1U_{t_1}A_2U_{t_1}^*U_{t_1}U_{t_2}\otimes l_{t_1}l_{t_2}=A_1A_2(t_1)U_{t_1+t_2}\otimes l_{t_1+t_2}
\end{multline}

We can now introduce the second crossed product.
\begin{satz}With $\mcal{R}(\mcal{M},\alpha)$ the first crossed product we define the second crossed product $\mcal{R}(\mcal{R}(\mcal{M},\hat{\alpha})$ with $\hat{\alpha}$ the representation of the dual group $\hat{\mbb{R}}$ on $\mcal{R}(\mcal{M},\alpha)$ via $W_p\mcal{R}(\mcal{M},\alpha)W_p^*$. It holds that  $\mcal{R}(\mcal{R}(\mcal{M},\hat{\alpha})$ is isomorphic to $\mcal{M}\overline{\otimes}\mcal{B}(L_2(\mbb{R}))$.
\end{satz}

While $\mcal{R}(\mcal{M},\alpha)$ acts on $\mcal{H}\otimes L_2(\mbb{R})$ and is generated by $\{A\otimes\mbf{1},U_t\otimes l_t\}$, the second crossed product acts on $L_2(\mbb{R},\mcal{H}) 
\otimes L_2(\mbb{R})$ or $\mcal{H}\otimes L_2(\mbb{R})\otimes  L_2(\mbb{R})$ and is generated by 
\beq \{A\otimes\mbf{1}\otimes\mbf{1},U_t\otimes l_t\otimes\mbf{1},\mbf{1}\otimes W_p\otimes w_p\} \eeq
(the long proof of the above results can be found in e.g. \cite{Kad2} p.970ff).
\begin{koro}If $\mcal{M}$ is a properly infinite v. Neumann algebra $\mcal{R}(\mcal{R}(\mcal{M},\hat{\alpha})$ is $\star$-isomorphic to $\mcal{M}$.
\end{koro}

The last point we want to discuss in this section is the case when the automorphism group is the modular group $\sigma_s$.
\begin{satz}If $\mcal{M}$ is countably decomposable or $\sigma$-finite (it is sufficient if $\mcal{H}$ is  separable) and of type $III$ and with $\sigma_s$ the modular group corresponding to a faithful normal state $\omega$, then $\mcal{R}(\mcal{M},\sigma)$ is of type $II_{\infty}$.
\end{satz}
(cf. \cite{Kad2} p. 974ff).
\section{The Physical Content of the Crossed Product Construction}
In this section we want to clarify which physical role the various constituents in the crossed product construction are playing, most notably the modular automorphism group, $\sigma_s$, implemented by the modular operator $\Delta^{is}$ and the  properties of the projectors in a type $III$ v. Neumann algebra. As in \cite{types} we think that there does exist a deep preastablished correspondence between the extremely rich mathematics in the field of v. Neumann algebras and the corresponding concepts of quantum physics. We think, in particular, that the analysis may yield a first step away from the traditional semiclassical picture of QFT in classical space-time into QG proper.
\subsection{The Role of the Modular Group}
There do exist a large number of papers about the mathematical implications of the socalled modular group and its role in Quantum Statistical Mechanics and QFT (see e.g. \cite{Haag}), but not so many concerning more fundamental properties in the microphysical field. This remark needs however some explanations. On the one side, its quasi automatic occurrence in the context of type $III$ v. Neumann algebras is intriguing. Furthermore, if one believes in the above mentioned preastablished harmony between physics and mathematics in this context, one would like to better understand its role in a more fundamental sense, as e.g. in QG. Such a better understanding is, for example, attempted in \cite{Connes},\cite{Rov1},\cite{Rov2},\cite{Dirac}. 
\begin{bem} In a few cases there exists a geometric role of the modular group like in the Rindler wedge case, but we think its true nature is a different one (see below).
\end{bem}

In \cite{Connes},\cite{Rov1},\cite{Rov2} the main interest is the role of the modular evolution in the context of the \tit{problem of time} in QG and the concept of  \tit{thermal time} made possible due to the existence of the modular structure. In \cite{Dirac} we attempted to derive the modular structure as a necessary consequence of the microscopic deep structure of the quantum vacuum or QST, most notably from the a priori existence of the fluctuation spectrum of the quantum (vacuum) fluctuations permanently roaming through QST.

In this context we want to mention the point of view of J.A.Wheeler, expressed in his talk in \cite{Wheeler}
\begin{quote}
Second, we accept as a tentative working hypothesis the picture of Clifford and Einstein that particles originate from geometry; that there is no such thing as a particle, immersed in geometry, but only a particle built out of geometry.
\end{quote}

In a general covariant theory there is no preferred time flow and the dynamics of the theory cannot be formulated in terms of an evolution in a single external parameter. It is the main message of \cite{Connes} to attribute the temporal properties of the time flow to thermodynamical causes. It is plausible to develop a statistical approach for systems, having a large number of degrees of freedom (DoF) which are only insufficiently controlled. A case in point is a QFT in a local region, $\mcal{O}$, or in a more fundamental layer, QG or QFT interacting with QST in a certain region $\mcal{O}$. In such cases a statistical approach seems to be natural.

Local observable algebras in QFT are typical of type $III$. Let $\Omega$ be a cyclic and separating vector, coming from a faithful normal state $\omega$, we define the operator:
\beq S:A\Omega\to A^*\Omega\quad ,\quad A\in \mcal{R}(\mcal{O})  \eeq
the algebra of local observables. The polar decomposition yields
\beq S=J\Delta^{1/2}  \eeq
with $J$ antiunitary  (the modular conjugation) and $\Delta$ a positive, in general unbounded operator. The Tomita-Takesaki theorem states that the map $\sigma_s$, defined by
\beq \sigma_s(A)=\Delta^{is}A\Delta^{-is}  \eeq
leaves $\mcal{R}(\mcal{O})$ invariant, the group $\sigma_s$ being called the modular automorphism group. With $\Delta:= \exp H_T$, $H_T$ is called the Tomita or modular Hamiltonian.
\begin{bem} Note that in general $H_T$ does not have the structure of an ordinary particle Hamiltonian, therefore its thermodynamic properties are not entirely obvious. A typical example are the Lorentz boosts in the Rindler wedges, occurring as modular evolution, which certainly do not have ordinary particle character.
\end{bem}

The modular group is called inner if there exists a unitary group $U(s)$ with $U(s)\in\mcal{M}$ (a v. Neumann algebra) and
\beq \sigma_s(A)=U(s)AU^*(s)  \eeq
Typically the modular group is not inner, but if there exists a unitary group $U(s)\in\mcal{M}$ such that 
\beq \sigma^2_s(A)=U(s)\sigma^1_s(A)U^*(s)  \eeq
for two modular groups on $\mcal{M}$, related to two different faithful normal states, $\omega^1,\omega^2$ on $\mcal{M}$, they are called inner equivalent.
\begin{satz}The co-cycle Radon-Nikodym theorem of Connes (\cite{Connes2}) states that on $\mcal{M}$ two modular groups $\sigma^1,\sigma^2$ given by $\omega^1,\omega^2$ are inner equivalent. The equivalence class defines the outer modular automorphism which is uniquely defined by $\mcal{M}$.
\end{satz}
In this context the following theorem is useful in exhibiting the difference between type $II_{\infty}$ and $III$ (see e.g. \cite{Sunder} p.88, the original source is \cite{Tomita}):
\begin{satz}The following conditions are equivalent:\\
i) $\mcal{M}$ is semifinite.\\
ii) $\sigma_s^w$ is inner for some faithful, normal, semifinite weight (fns) $w$ on $\mcal{M}$.\\
iii) $\sigma_s$ is inner for every faithful, normal, semifinite weight on $\mcal{M}$.
\end{satz}
\begin{bem}Note that semifiniteness implies the existence of a fns trace. That is, the results of Connes show that there exists a trivial flow on $\mcal{M}$ and hence all flows are inner.
\end{bem}
\begin{koro}It follows that the Tomita action on a type $III$ algebra is outer. I.e., the spectral projections of $H_T$ do not lie in the original $\mcal{M}$.
\end{koro}

In \cite{Connes}the thermal origin of the modular action is emphasized. From a different starting point we came to a similar result in \cite{Requ2} concerning the thermal substructure of GR (this point of view has a larger history which is sketched in \cite{Requ2}). One starting point is the general observation that a system, consisting of many DoF which are or, equivalently, can by necessity only be insufficiently controlled, should be treated as a statistical or even thermal system.

In this context we now want to formulate a conjecture how the thermal time concept of Connes-Rovelli may be related to an important consequence of GR, i.e., the dependence of the rate of an ideal clock in a gravitational field. With
\beq ds^2=g_{ik}(x)dx^idx^k    \eeq
and $dx^0=dt$, the coordinate time, we assume the clock occupies the fixed spatial position $x^i,i=1,2,3$. The \tit{proper time}, according to GR, is then
\beq d\tau^2=|g_{00}|(x)dt^2  \eeq
Take e.g. the example of a static BH with Schwarzschild coordinates. we have
\beq d\tau^2=(1-2M/r)^2dt^2  \eeq
We see that coordinate time, $dt$, is displayed by a clock at radial distance $r\to\infty$ and that the clock near the event horizon goes slower.

We now formulate the following conjecture:
\begin{conjecture}
The microscopic cause for the rate dependence of a clock in a gravitational field is the interaction of the clock with the deformed quantum fluctuation spectrum induced by the Tomita or modular Hamiltonian $H_T$. Furthermore, we conjecture that the local time at $x$ is given by the evolution of the local modular group, $\Delta^{is}=\exp isH_T$.
\end{conjecture}
That is, we think that this is a typical example exhibiting the interaction of quantum gravity DoF with the Tomita Hamiltonian $H_T$ and showing how gravity is making visible its presence in the microscopic quantum underground in QST. 

This feature can be further corroborated by the following interesting and important educated speculation by Sakharov and some other colleagues (emergent gravity, an incomplete list is \cite{Sak},\cite{Visser},\cite{Adler}). 
\begin{conjecture}(Sakharov) General Relativity arises as an emergent property of the vacuum fluctuations of matter fields.
\end{conjecture}
In our approach we observe that the local spectrum of quantum fluctuations is strongly distorted by gravity. We can go one step further and say:
\begin{assumption}The local spectrum of quantum fluctuations represents, a fortiori, the gravitational field or the local microscopic state of space-time (or QST). This picture then neatly corresponds with the ideas of Sakharov.
\end{assumption}

In \cite{Dirac} we made a detailed analysis of the consequences of the existence of the modular action for the form of the excitation patterns in the quantum vacuum. A fortiori, we argued that, conversely, the particular structure of the excitation patterns in the quantum vacuum implies the emergence of the modular structure of local v. Neumann algebras. Among other things we argued in favor of the old Dirac picture, that is, a densely filled Dirac sea of excitations. Invoking the ingeneous Landau picture of low lying elementary collective excitations (see e.g. \cite{Landau1},\cite{Landau2}) we assume that approximately we can treat the quantum vacuum as a thermal system with a constant creation and annihilation of extra additional exciation modes or holes in the Dirac sea. We argue that this is the scenario which is seen within a local v. Neumann algebra.
\begin{bem}The crucial idea of Landau was it to replace a complicated strongly interacting system of many DoF by a relatively weakly interacting system of collective excitations.
\end{bem} 

In \cite{Dirac} we provided both an approximative but more physical picture and a rigorous analysis of the situation using the full machinery of the modular theory (for more details see \cite{Narnhofer} and \cite{Requ3}). In the following we restrict ourselves with invoking only the physical picture which is however only approximately correct. We assume the existence of creation and annihilation operators for certain collective excitation modes in the quantum vacuum. We call this the real or observable process. But our main point is that such a real process is a temperature dependent superposition of, for example, the annihilation of an excitation lying above the socalled Fermi surface (given by the zero energy level of the quantum vacuum $\Omega$) and the creation of a corresponding hole excitation in the Dirac sea and vice versa. We note that a similar picture was discussed in \cite{Umezawa} in the case of a finite volume Gibbs state, see also \cite{Ojima2} (without a physical interpretation such a representation does already occur in \cite{Araki}).      

We now conjecture that the real space creation and annihilation operators can be written in the following form. If we have translation invariance, $k$ denotes the momentum. In the local algebras, where we do not have translation invariance, it labels the members of the canonical commutation relations.
\begin{conjecture} We call the operators $a^{(+)}(k)$ real-space variables, the (temperature dependent) operators
$a^{(+)}(k,\beta),\tilde{a}^{(+)}(k,\beta)$ primordial variables. 
 \beq a(k)=\cosh c_k\cdot a(k,\beta)+\sinh c_k\cdot\tilde{a}^+(k,\beta)   \eeq
where it is assumed that $a(k),a(k,\beta),\tilde{a}(k,\beta)$ and their adjoints satisfy the same (bosonic) canonical commutation relations, e.g.:
\beq [a(k),a^+(l)]=\delta (k-l)   \eeq
$a(k,\beta)$ annihilates a mode above the Fermi surface, $\tilde{a}(k,\beta)$ annihilates a hole below the Fermi surface, that is, $\tilde{a}^+(k,\beta)$ creates a hole of energy-momentum $(-\omega_k,-k)$. This implies that the tilde operators commute with the $a^{(+)}(k,\beta)$.
\end{conjecture}
\begin{bem} In the following $\omega_k$ is chosen positive.
\end{bem}
The following is an important observation. It shows that in the same system does exist a representation which commutes with the preceding one.
As the $a^{(+)}(k,\beta),\tilde{a}^{(+)}(k,\beta)$ operators are assumed to fulfill the canonical commutation relations separately, we can construct another representation in \tit{real space}, i.e., we define:
\beq \tilde{a}(k):=\cosh c_k\cdot \tilde{a}(k,\beta)+\sinh c_k\cdot a^+(k,\beta)   \eeq
\begin{ob} This tilde representation again commutes with the $a^{(+)}$-representation.
\end{ob}
\begin{bem} The coefficients $c_k$ are also functions of $\beta$ (see below). 
\end{bem}

The backtransform now reads:
\beq a(k,\beta)=a(k)\cdot\cosh c_k-\tilde{a}^+(k)\cdot\sinh c_k   \eeq
\beq \tilde{a}(k,\beta)=\tilde{a}(k)\cdot\cosh c_k-a^+(k)\cdot\sinh c_k  \eeq
\begin{bem} Similar relations hold for fermions.
\end{bem}
\begin{ob} In the primordial variables we have the following Hamilton operator:
\beq H_0=\int d^3k\, \omega(k,\beta)\cdot (a^+(k,\beta)a(k,\beta)-\tilde{a}^+(k,\beta)\tilde{a}(k,\beta))   \eeq
which reads in the real space variables:
\beq H_0=\int d^3k\, \omega(k,\beta)\cdot (a^+(k)a(k)-\tilde{a}^+(k)\tilde{a}(k))  \eeq
\end{ob}
\begin{bem} We see that via the finer analysis with the help of the primordial variables we get the kind of doubling we found in the Tomita formalism.
\end{bem}

\begin{postulate} The equilibrium state (a Hilbert space vector) $|0,\beta>$ is annihilated by $a(k,\beta),\tilde{a}(k,\beta)$:
\beq a(k,\beta)|0,\beta>=0=\tilde{a}(k,\beta)|0,\beta>   \eeq
It is hence not annihilated by the $a(k),\tilde{a}(k)$.
\end{postulate}
\begin{bem} Note the simlarity to the Unruh-black hole-situation.
\end{bem}
\begin{ob} After some calculations (see for example \cite{Umezawa}) we get:
\beq (\sinh c_k)^2=(e^{\omega(k)}-1)^{-1}=:f_B(\omega(k))   \eeq
That is, we have:
 \beq a(k)=(1+f_B)^{1/2}\cdot a(k,\beta)+(f_B)^{1/2}\cdot\tilde{a}^+(k,\beta)  \eeq
 \beq\tilde{a}(k)=(1+f_B)^{1/2}\cdot \tilde{a}(k,\beta)+(f_B)^{1/2}\cdot a^+(k,\beta)  \eeq
\end{ob}

Concluding our above arguments we show that the KMS-condition supports in fact our picture of quasi particle excitation modes and corrsponding holes in a Dirac sea. This was already exploited by us in e.g. \cite{Narnhofer}. The KMS-property yields for the Fourier transform of the two-point function of the type $(\Omega|A(x,t)B\Omega)$
\beq Re\, J(-k,-\omega)=e^{-\beta\omega}\cdot Re\, J(k,\omega)\; ,\;  Im\, J(-k,-\omega)=-e^{-\beta\omega}\cdot Im\, J(k,\omega)  \eeq

In this subsection we advanced mainly the argument of a thermal character of the modular action and the Tomita Hamiltonian. We said already above that a geometric action is presumably not generic for the Tomita evolution. However, as it turns out to be of tantamount importance to arrive at a better understanding of the concrete action of the modular group, there are various attempts to perturb or generalize, so to speak, some given geometric action, in order to arrive at new modular groups which are perhaps no longer of a geometrical type. This, however, turns out to be quite intricate. A very incomplete list is \cite{Bo},\cite{Schroer},\cite{Saffary}, the latter reference being a nice report of the overall situation at that time.
\subsection{The Role of the Class of admissible Projectors and Procedures in the Transition from Type $III$ to Type $II_{\infty}$}
In the following we want to discuss the subtle role the class of admissible projectors and procedures are playing in the transition of our local v. Neumann factors from type $III$ to $II_{\infty}$. We begin with a simple observation:
\begin{bem}Given two closed subspaces $P\mcal{H},Q\mcal{H}$, both with countably infinite dimension and with $ON$-bases $\{e_i\},\{f_i\}$, then it is easy to construct a partial isometry $V$ by defining 
\beq V:\; e_i\to f_i\; \text{and linearly extended with}\; V(\mbf{1}-P)\mcal{H}:=0  \eeq
\end{bem}
That is, the crucial point is rather, under what conditions $V$ does belong to $\mcal{M}$. We hence have to clarify the particular properties the admissible procedures, incorporated by $V$, have to fulfill.

To approach this question we formulate the following conjecture:
\begin{conjecture}Following our working philosophy of a parallelism of mathematics and physics in this field of the role of v. Neumann algebras, we assume that all operators, belonging to $\mcal{M}$ (a local v. Neumann factor), have a physical meaning and play a physical role in the quantum regime. This is obvious concerning the class of projectors, but we think, it holds as well concerning the class of partial isometries.
\end{conjecture}

Without going into the quantitative details we associate the notion of \tit{microscopic information} to a a state in $\mcal{H}$. If, for example, $\psi$ does not lie in $P\mcal{H}$, we say that $P$ applied to $\psi$ cuts off some or reduces the information content of the state in the process. We call such a process  irreversible. Note that in general a measurement is accompanied by a macroscopic irreversible (avalanche) process.
\begin{defi}
 On the other hand, the partial isometry $V\in\mcal{M}$ with $P\mcal{H}$ the initial space and $Q\mcal{H}$ the final space ($VV^*=Q,V^*V=P$)
\beq V:P\psi\to Q\psi   \eeq
is said to deform or redistribute the microscopic information but does conserve it. With  $V^*$ being the inverse of $V$ on $Q\mcal{H}$:
\beq V^*:Q\psi\to P\psi\;, \; V^*VP\psi=P\psi  \eeq
we call the process $V,V^*$ reversible between $P\mcal{H},Q\mcal{H}$.
\end{defi}

Interesting in such a context is the type $III$ case. In this situation we have that all projectors turn out to be equivalent.
\beq P\sim Q\; ,\; \forall\, P,Q\in\mcal{M} \;\text{and in particular}\; P\sim\mbf{1}  \eeq
i.e., using our above notion:
\begin{ob}In the type $III$ case all partial isometries in $\mcal{M}$ preserve the microscopic information.
\end{ob} 
More specifically we have
\begin{propo}If $\mbf{1}\sim P$ with $V:\mcal{H}\to P\mcal{H}$ we have that $\mcal{M}$ and $V\mcal{M}V^*$ are $\star$-isomorphic and, a fortiori, unitarily equivalent with Hilbert spaces $\mcal{H},P\mcal{H}$. By the same token, spectral measures of $A\in\mcal{M}\,,\,VAV^*\in
 V\mcal{M}V^*$ are unitarily equivalent and their spectra are the same, i.e.
 \beq E_A(S)\to VE_A(S)V^*  \eeq
 \end{propo}
Proof: We have for example
\beq  (V\psi|VAV^*V\psi)=(V^*V\psi|AV^*V\psi)=(\psi|A\psi)  \eeq
as $V^*V=\mbf{1}$. Furthermore, we can compare $P\mcal{M}P$ with $V\mcal{M}V^*$. We have
\beq P\mcal{M}P=VV^*\mcal{M}VV^*=V(V^*\mcal{M}V)V^*\subset V\mcal{M}V^*  \eeq
But with the choice $VAV^*\,,\, A\in\mcal{M}$ on the rhs we get
\beq V^*VAV^*V=A  \eeq
i.e., the map $VAV^*$ is one to one. 
\begin{koro} We have $P\mcal{M}P=V\mcal{M}V^*$.
\end{koro}
\begin{bem} Corresponding results hold for the case $P\sim Q$ and $V:P\mcal{H}\to Q\mcal{H}$
\end{bem}
Some remarks on the type $III$ situation can, for example, also be found in \cite{Yngvason}.

To give a more intuitive picture of the above ideas, we mention e.g. the Stern Gerlach experiment. An ingoing up- and down superposition of z-spin passes through an inhomogeneous magnetic field. The ingoing state, $\psi$, is split (or rather, separated) into two components
\beq  \psi\to\psi_1+\psi_2  \eeq
but it is still a superposition, not a mixture. One may say, it is at this point a \tit{premeasurement}. Note that, in principle, one can reunite the superposition $\psi_1+\psi_2$ to $\psi$ by using an appropriate magnetic field. One may now associate the process
\beq \psi\to\psi_1+\psi_2 \;\text{to}\;V:\psi\to\psi_1+\psi_2  \eeq
and the reverse process
\beq \psi_1+\psi_2\to \psi \;\text{to}\; V^*:\psi_1+\psi_2\to\psi  \eeq
\begin{conclusion}We think, in the type $III$ case the partial isometries $V,V^*$ are working in just this sense as illustrated in the above example.
\end{conclusion}

This is no longer the case in the type $II_{\infty}$ situation. In that case there exist subspaces, $P\mcal{H}$, which cannot reversibly be mapped by a partial isometry, $V^*$, isometrically onto $\mcal{H}$. That is, in this case microscopic information in the sense we defined it above, is lost.

We now want to study the transition, implemented by the crossed product construction, from type $III$ to type $II_{\infty}$. For us, the interesting case is the modular automorphism group, $\sigma_s$, and its implementation by $\Delta^{is}=\exp isH_T$. The crossed product, $\mcal{R}(\mcal{M},\sigma_s)$, is generated by $\{A\otimes\mbf{1},\Delta^{is}\otimes l_s\}$ with $A\in\mcal{M},l_s=\exp -is\hat{p}$, $\hat{p}$ the momentum operator on $L_2(\mbb{R})$.

In a first step we have to study which kind of projection operators do occur in $\mcal{R}(\mcal{M},\sigma_s)$ compared to $\mcal{M}$. The projectors, $\{P\}$, lying in $\mcal{M}$, go over into $\{P\otimes\mbf{1}\}$. As $\exp isH_T\otimes\mbf{1},\mbf{1}\otimes \exp -is\hat{p}$ do commute, we have, furthermore, projectors of the kind $E_{\Delta}\otimes e$, $E_{\Delta}$ a spectral projector of $H_T$, $e$ of $\hat{p}$.
\begin{conclusion}The projectors of $\mcal{M}$ and of $H_T\otimes\hat{p}$ go over, more or less unchanged into the corresponding projectors of  $\mcal{R}(\mcal{M},\sigma_s)$.
\end{conclusion}

Much more interesting are the projectors which arise by a nontrivial combination of members $A_j\otimes \mbf{1}$ with $A_j\in\mcal{M}$ and $\Delta^{is_j}\otimes l_{s_j}$. We remind the reader of the particular product rules, proved at the end of section 2, which show that the members of 
 $\mcal{R}(\mcal{M},\sigma_s)$ consist of sums of form invariant terms $A_j\cdot\Delta^i{s_j}\otimes l_{s_j}$. It is easy to construct s.a. elements in  $\mcal{R}(\mcal{M},\sigma_s)$ by forming sums, $(\hat{B}+\hat{B}^*)$ with $\hat{B}\in \mcal{R}(\mcal{M},\sigma_s)$. We are interested in the spectral projections of such s.a. operators. More specifically, we are interested how they are constructed from the above generators $A_j\otimes\mbf{1}$ and  $\Delta^{is_j}\otimes l_{s_j}$.

We begin with the class of finite polynomials
\beq \sum_{j=0}^n\, c_j\hat{B}^j\;,\, \hat{B}\;\text{a s.a. element of}\;\mcal{R}(\mcal{M},\sigma_s) \eeq
As $\hat{B}$ is bounded, its spectrum, $Spec(\hat{B})$, lies in a finite closed interval $I=[-||\hat{B}||,||\hat{B}||]\subset \mbb{R}$. The Stone-Weierstrass theorem says that the set of finite polynomials is dense in the set of continuous functions
\beq f(x)\in \mcal{C}(I)\;\text{with norm}\; f_{\infty}(x)= \sup f(x)  \eeq

With $dE(\lambda)$ the spectral meaasure of $\hat{B}$ we have hence
\begin{lemma}The finite polynomials, $P(\hat{B})$, are dense in the set $f(\hat{B})=\int\, f(\lambda)dE_{\lambda}$, $f$ continuous in $I$, in the norm $||A||$.
\end{lemma}
Projectors are given by characteristic functions, $I_{\Delta}$, $\Delta$ an interval covering part of the spectrum of $\hat{B}$.

Employing the measure class given by $dE_{\lambda}$, we can use the Lebesgue dominated convergence theorem to approximate $I_{\Delta}$ by a sequence of functions from $\mcal{C}(I)$.
\begin{conclusion}We see that spectral projectors of$\hat{B}$ can be approximated by (infinite) sequences of polynomials $P(\hat{B})$. That is, they consist of many terms of products of $A_j\otimes\mbf{1}$ and $\Delta^{is_j}\otimes l_{s_j}$. It follows that in the construction of such projectors, elements of $\mcal{M}$ and the modular group, $\Delta^{is}$ are typically combined in a very complicated way.
\end{conclusion}
\subsection{Conclusion}
It is our aim to understand to what extent the properties of projectors are affected by the coupling of $\mcal{M}$ with the Tomita evolution, $\exp isH_T$, leading to a larger algebra of observables, $\mcal{R}(\mcal{M},\sigma_s)$. We described the physical content of partial isometries , $V:P\mcal{H}\to Q\mcal{H}$, by saying that this process preserves the total microscopic information content but redistributes it. As an example of what we have in mind we provided the Stern-Gerlach experiment.

We said that, while $\sigma_s$ acts on $\mcal{M}$, for example the spectral projections of $H_T$ are in general not observable if $\sigma_s$ is an outer automorphism. We argued that this changes if $\mcal{M}$ and $\exp isH_T$ are really coupled as in $\mcal{R}(\mcal{M},\sigma_s)$.

We argued above that $\exp isH_T$ acts on the quantum fluctuation spectrum and, a fortiori, may even be related to quantum gravity effects. Following our working philosophy that the class of projectors and the admissible partial isometries encode the various degrees of microscopic information, contained in them, we make the following claim:
\begin{conjecture}
While we will not prove it mathematically, we argue on physical grounds (we have explained above) that the spectral projectors, belonging to typical observables $\hat{B}$ of  $\mcal{R}(\mcal{M},\sigma_s)$ and which are formed via a complicated coupling of elements of $\mcal{M}$ and $\exp isH_T$ are not related to projectors lying in, for example, $\mcal{M}\otimes \mbf{1}$, or the identity projector $\mbf{1}$ via partial isometries as their information content are different. Hence $\mcal{R}(\mcal{M},\sigma_s)$ is no longer of type $III$.
\end{conjecture}
\begin{bem}This holds the more so as quantum gravity effects may play a role in this context as we have argued above in connection with the ideas of Sakharov. We learned from our previous discussions that the fluctuation spectrum is strongly affected by gravity.
\end{bem}

\end{document}